\documentclass[12pt]{article}

\begin{document}

\def\beq{\begin{equation}}
\def\eeq{\end{equation}}

\begin{titlepage}

\begin{flushright}
Padua University\\
December 1998\\
\end{flushright}
\vskip2truecm

\begin{center}

{ \Large Action for $IIB$ Supergravity in 10 dimensions}

\vspace{1cm}

{ Gianguido Dall'Agata$^{\dag}$}\footnote{dallagat@to.infn.it}, { 
Kurt 
Lechner$^{\ddag}$}\footnote{kurt.lechner@pd.infn.it},  and
{Mario Tonin$^\ddag$}\footnote{mario.tonin@pd.infn.it} 
\vspace{1cm}

{$\dag$ \it Dipartimento di Fisica Teorica, Universit\`a degli 
studi di Torino, \\
 via P. Giuria 1,I-10129 Torino}

\medskip

{ $\ddag$ \it Dipartimento di Fisica, Universit\`a degli Studi di 
Padova,

\smallskip

and

\smallskip

Istituto Nazionale di Fisica Nucleare, Sezione di Padova, 

Via F. Marzolo, 8, 35131 Padova, Italia
}

\vspace{1cm}

\begin{abstract}
We review the construction of a manifestly covariant, supersymmetric and 
$SL(2R)$ invariant action for $IIB$ supergravity in D=10.
\end{abstract}

\end{center}
\vskip 0.5truecm
\noindent Talk given at the workshop {\it Quantum Aspects of Gauge 
Theories, Supersymmetry and Unification}, Greece, September 1998.

\end{titlepage}

\newpage

\section{Introduction}

$D=10$, $IIB$ Supergravity acquired new life, on one hand from the 
discovery of different dualities among string theories and among 
these  and $D=11$ supergravity and, on the other hand, from the related 
discovery of $D$-branes. Its bosonic sector consists of the graviton, 
two scalars, a complex rank two tensor and a real rank four tensor with 
selfdual field strength. The fermions are two gravitinos with the same 
chirality and two spin 1/2 fermions with opposite chirality w.r.t.
the gravitinos. The presence of a self--dual tensor (chiral boson) 
is at the origin of the difficulty related with a covariant lagrangian 
formulation of this theory.
A common statement is that such a formulation cannot exist.
The statement is correct if with "covariant 
action" one means a "globally defined covariant action", but sometimes it 
can be useful to deal with covariant actions even if they are not 
globally defined \footnote{For instance, in a recent paper \cite{8} the 
covariant action of type $IIB$ supergravity obtained in \cite{3},
\cite{4} has been 
used to obtain the quadratic action for the physical fields in an $ 
AdS_5 \times S_5$ background.}. 

Recently a new method \cite{1}, \cite{2} has been proposed to write  
covariant actions with manifest duality and/or in presence of chiral 
bosons and it is natural to apply this method to get a manifestly 
covariant action for $D=10$, $IIB$ supergravity. This action has been 
constructed in \cite{3}, at the bosonic level, and in \cite{4} 
for the complete theory, and will 
be presented in section 3 \footnote{For a different approach, see 
\cite{9}.}.
To write the complete action, one needs the trasformation rules 
under supersymmetry. The supersymmetry 
trasformations and the field equations of $IIB$ supergravity are well 
known at a non lagrangian level \cite{5}, \cite{6} and can be conveniently 
derived in a superspace approach \cite{4}, \cite{6}, \cite{7}. 
In section 2 we review and sligthly improve this approach, the 
improvement being that we present the rheonomic parametrizations not only 
for the basic curvatures but also for their duals.

\section{Superspace approach}

In $D=10$, $IIB$ supergravity the superspace coordinates are $Z^{M} \equiv 
(x^m,\theta^\mu,\bar{\theta^\mu})$,  
 where $x^m$ ($m=0,...,9$) are the space--time 
coordinates and $\theta^\mu$ ($\mu=1,...,16$) with their complex conjugates 
$\bar{\theta^\mu}$ are Grassmann variables. Our normalization for a 
$p$-superform $Y_p$ is $Y_p = 1/p! \, dZ^{M_p}...dZ^{M_1} \, Y_{{M_1}...{M_p}}$ 
where the 
$dx^m$ are anticommuting and the $d\theta^\mu$ are commuting, so that 
we can suppress the wedge symbol in the products of (super)forms.
$D=10$, $IIB$ supergravity is invariant under the global S duality group
$SL(2,R) \approx SU(1,1)$, which acts only on the bosons. To keep this 
symmetry manifest, the scalars are described by two complex 
$SU(1,1)$ doublets $V_\pm^I$, ($I=1,2$), 
which parametrize the coset $SU(1,1) \over U(1)$ and satisfy 
$V_{+I} V_-^I=1$.
The indices $I,J$ are lowered and rised from the left with the metric 
${\epsilon}_{IJ}$ and ${\epsilon}^{IJ}$ where ${\epsilon}_{12} = 
- {\epsilon}_{21} = 1 = {\epsilon}^{21} = -{\epsilon}^{12}$. 
$V_+^I$ and $V_-^I$ are superfields (0-superforms) with 
$U(1)$--charges $+1$ and $-1$ respectively and satisfy 
the reality condition $V_-^I = \overline{V_+^I}$
\footnote{The complex conjugate 
$\phi^*$ of a complex field $\phi$ will be denoted by $\bar\phi$ but for 
an $SU(1,1)$ doublet $\psi = \quad{ \psi_1 \choose \psi_2} \quad$ we define
 $\bar{ \psi} = \tau_1 \psi^*$ where $\tau_1 = \quad { 0\> 1 \choose 1\> 
0}\quad $.}.
The $U(1)$ connection is the one-superform $Q = i V_+^I 
dV_{-I}$ and the curvatures of the scalars are 
\beq
\label{1}
R_1 = V_{+I} dV^I_{+},\quad \bar{R}_1 = V_-^I \, dV_{-I}, 
\eeq
which are $SU(1,1)$ invariant one-superforms with $U(1)$ charges $\pm 2$.
The other geometrical objects are the supervielbeins $e^A$, the Lorentz 
valued superconnection ${\omega_A}^B$, a complex antichiral spinor 
superfield $\Lambda$ and the $p$--superform potentials $A_p$, where 
$p=2,4,6,8$. 
The supervielbeins $e^A \equiv (e^a,e^\alpha,\bar{e}^\alpha)$, with torsion
 $T^A = De^A$ ($a=0,...,9;$ $\alpha = 1,...,16$), describe the graviton and 
the gravitinos. 
The superconnection ${\omega_A}^B$, with Lorentz 
curvature ${R_A}^B$, can be expressed in terms of $e^A$ due to 
the torsion constraints. 
The gravitinos $e^{\alpha} \equiv \psi^{\alpha}$
and $\bar {e}^{\alpha} \equiv \bar{\psi}^\alpha $ are complex conjugate
chiral spinor one-superforms.  $\Lambda$ and $\bar{\Lambda}$ describe the 
spin 1/2 fermions. 
  $ D = d + \omega + iqQ $ denotes the $U(1)$ and $SO(1,9)$ covariant 
differential for a superfield or a superform with $U(1)$ charge $q$.
The two rank--two tensors are described by the $SU(1,1)$ 
doublet two-superform $A_2^I = - \bar{A}_2^I$, and the chiral boson by the 
real four-superform $A_4$.
In addition to $V_\pm^I$ 
only $e^{\alpha}$ and $\Lambda_{\alpha}$ are charged under
$U(1)$ and have charges 1/2 and 3/2 respectively (and, of course 
$\bar{e}^{\alpha}$ and $\bar{\Lambda}_{\alpha}$ have charge -1/2 and 
-3/2).
We shall also introduce the duals of $A_2$, that is, the  doublet 
six-superform $A_6^I = \bar{A}_6^I$, and the duals of the scalars.
Here we meet a little surprise: the duals of the scalars are a triplet of 
eigth-superforms ${A_{8I}}^J = \bar{A}_{8I}{}^J$ which belong to the adjoint 
representation
of $SU(1,1)$. The space-time differentials ${dA_{8I}}^{J} |_{\theta=0}$ are the 
Hodge duals of the three $SU(1,1)$ conserved currents; they are closed 
and, therefore, locally exact. However, they are not independent. There is 
a linear relation between them so that, after the gauge fixing of the 
local $U(1)$ simmetry, only two scalars and two dual eigth--forms survive.
The curvatures for the superform potentials $A_p$ are
\beq 
R_3^I = dA_2^I, \label{2} 
\eeq
\beq 
R_5 = dA_4 +iA_{2I} \, dA_2^I, \label{3}
\eeq
\beq
R_7^I = dA_6^I -
i dA_4\, A_2^I +1/3 \, A_{2J} \, dA_2^J \, A_2^I, \label{4}
\eeq
\beq
R_9^{(IJ)} = dA_8^{(IJ)} -
1/2 \, \left((dA_6^I + i/2 \, dA_4 \, A_2^I - 1/12 \, A_2^I 
A_{2K} \, dA_{2}^{K}) \, A_2^J + I 
\longleftrightarrow J\right),  \label{5} 
\eeq 
and the gauge transformations of the potentials are such 
that these curvatures are invariant.

Instead of $R_3^I$, $R_7^I$, ${R_{9J}}^I$ it is convenient to use the 
$SU(1,1)$ invariant curvatures 
\beq 
R_3 = - V_+^IR_{3I}, \quad \bar{R}_3 = - 
V_-^IR_{3I}, \quad R_7 = - V_+^IR_{7I}, \quad\bar{R}_7 = 
V_-^IR_{7I}, 
\label{6}
\eeq
\beq 
R_9 = V_+^I \, {R_{9I}}^J \, V_{+J},\quad \bar{R}_9 = V_-^I \, 
{R_{9I}}^J \, V_{-J}, \quad R_9^{(0)} = - V_+^I \, {R_{9I}}^J \, V_{-J}. 
\label{7}
\eeq
Taking the differential (or covariant differential) of these curvatures 
and of the torsion $T^A$ one can derive the associated Bianchi 
identities. Then, solving the Bianchi identities, under suitable 
constraints, one obtains the torsion and curvature rheonomic 
parametrizations (that include the constraints). For $R_p$, 
$p=1,3,5,7,9,9^{(0)}$ one gets 
\beq 
R_p = F_p - C_p, 
\label{8}
\eeq
where $F_p$ indicates the components of $R_p$ along $e^a$,  
i.e. 
\beq 
F_p=\frac{1}{p!} \, e^{a_1}...e^{a_p} \, F_{{a_p}...{a_1}}, \label{9}
\eeq 
and the $C_p$, which involve the gravitino superforms $\psi^\alpha$, 
$\bar{\psi}^\alpha$ and the spinor superfields $\Lambda_\alpha$ and 
$\bar{\Lambda}_\alpha$, are given by 
\beq 
C_1=2\psi\Lambda, 
\label{10}
\eeq
\beq 
C_3 = \frac{1}{2} \, e^a e^b \, 
(\bar{\psi}\Gamma_{ab}\Lambda) 
+\frac{i}{2} 
e^a \, (\psi\Gamma_a\psi) = C_3^\Lambda + C_3^\psi,  
\label{11}
\eeq  
\beq 
C_5 = \frac{1}{5!} \, e^{a_1}...e^{a_5}(\bar{\Lambda}\Gamma_{{a_1}...{a_5}}\Lambda) 
- \frac{1}{3!} \, e^a e^b e^c \, (\bar{\psi}\Gamma_{abc}\psi) = C_5^\Lambda + 
C_5^\psi,
\label{12}
\eeq
\beq 
C_7 = \frac{1}{6!} \, e^{a_1}...e^{a_6} \, (\bar{\psi}\Gamma_{{a_1}...{a_6}}\Lambda) - 
\frac{i}{2 \cdot 5!} \, e^{a_1}...e^{a_5} \, (\psi\Gamma_{{a_1}...{a_5}}\psi) = 
C_7^\Lambda + C_7^\psi,
\label{13} 
\eeq 
\beq
C_9 = \frac{2}{8!}e^{a_1}...e^{a_8} \, (\psi\Gamma_{{a_1}.
..{a_8}}\Lambda),
\label{14}
\eeq 
\beq
\begin{array}{rcl}
C_9^{(0)} &=& \displaystyle -i \, \left[\frac{3}{9!} \, 
e^{a_1}... e^{a_9}\, (\bar{\Lambda} \Gamma_{{a_1}... {a_9}}\Lambda)+ 
\frac{1}{2\cdot 7!} \, e^{a_1} ... e^{a_7} \, 
(\bar{\psi}\Gamma_{{a_1}...{a_7}}\psi)\right] 
\\
&=& 
C_9^{(0)\Lambda} + C_9^{(0)\psi}.
\end{array}
\label{15}
\eeq 
Here $C_p^\Lambda$ ($C_p^\psi$) 
indicates the terms of $C_p$ that depend on (are independent off) 
$\Lambda$.
For the torsion $T^A$ and for $D\Lambda$, the rheonomic 
parametrizations are 
\begin{eqnarray} 
T^a &=& De^a = i\bar{\psi}\Gamma^a\psi \\
T = D \psi &=& \displaystyle 
\frac{1}{2} \, e^a e^b T_{ba} 
+e^a \, \left[N_a\bar{\psi} + i\left(\frac{1}{192} \,
F_{a{b_1}...{b_4}}^{(+)}\Gamma^{{b_1}..{b_4}} + L_a\right)\psi\right] 
\nonumber\\
&-& \displaystyle
(\bar{\psi}\Lambda)\bar{\psi} - \frac{1}{2} (\bar{\psi}\Gamma^a\bar{\psi}) 
\Gamma_a
\Lambda \\
D\Lambda &=& e^b D_b\Lambda+\frac{i}{2} \, F^a\Gamma_a\bar{\psi}+ \frac{i}{24}
F^{abc}\Gamma_{abc}\psi,
\end{eqnarray}
where 
\beq 
L^a = -\frac{21}{2} \Lambda^a 
+\frac{3}{2}\Lambda^b\Gamma_{ab} + \frac{5}{4} \Lambda_{abc}\Gamma^{bc}-
\frac{1}{4} 
\Lambda^{bcd}\Gamma_{abcd}- 
\frac{1}{48}\Lambda_{a{b_1}..{b_4}}\Gamma^{{b_1}..{b_4}},
\eeq 
\beq 
N^a=\frac{3}{16}\left(-F_{abc}\Gamma^{bc} +\frac{1}{9}F^{bcd}\Gamma_{abcd}
\right),
\eeq 
$\Lambda_{{a_1}...{a_n}} = 
1/16 \, \bar{\Lambda}\Gamma_{{a_1}...{a_n}}\Lambda$, and 
$F_{{a_1}...{a_5}}^{(+)}$ is the self--dual part of $F_{{a_1}...{a_5}}$.

These parametrizations determine, on  one hand, the supersymmetry 
trasformations of the components fields, which are given by the 
covariant Lie derivatives  of the 
associated superfields along the susy parameter 
$\epsilon^A=(0,\epsilon^\alpha,\bar{\epsilon}^\alpha )$, 
evaluated at $\theta=d\theta=0$ (modulo 
$\epsilon$-dependent gauge trasformations).
Notice that $F_{{a_1}...{a_5}}^{(+)}$ enters only in the susy 
transformation of the gravitinos.

On the other hand, they force the theory to be on shell and imply, 
therefore,  
the field equations. The field equation for the rank--four tensor is 
just the self--duality condition for $F_5$. The equations of motion
for the rank-two 
tensors and for the scalars are equivalent to the duality conditions between
 $F_3$ and $F_7$ and between $F_1$ and $F_9$ respectively:
\beq 
*F_5=F_5,
\label{16}
\eeq 
\beq 
*F_3 = F_7, 
\label{17}
\eeq
\beq
*F_1 = F_9, 
\label{18}
\eeq
\beq 
0 = F_9^{(0)}. 
\label{19}
\eeq  
The hodge dual of $F_p$ is the $(10-p)$--superform
$(*F)_{10-p}$, whose intrinsic components are 
\beq
(*F)_{{a_1}...{a_{10-p}}} = 
\frac{1}{p!} \, {\epsilon_{{a_1}...{a_{10-p}}}}^{{b_1}...{b_p}}
F_{{b_1}...{b_p}}.
\eeq

Eq. (\ref{16}) is, indeed, the self-duality condition for the chiral tensor, 
and 
the field equations for the rank two tensors and the scalars are obtained 
by taking the covariant differentials of eqs. (\ref{17}) and (\ref{18}) and 
using the 
Bianchi identities for $R_7$ and $R_9$ respectively. Eq. (\ref{19}) is the 
linear relation between the three currents $d{A_{8I}}^J $, mentioned above,
since $F^{(0)}_9$ is purely immaginary.

\section{Covariant action}

In this section we will apply the general method of \cite{1}, \cite{2} 
to write a 
covariant, supersymmetric action for $D=10$, $IIB$ supergravity. Being a 
component action,  we shall deal with fields and forms, not 
superfields and superforms. 
However, we shall use the same symbols of the 
previous section to indicate those objects evaluated at 
$\theta=d\theta=0$. 

As a first step, let us discuss the action for a free 
chiral boson in a ten dimensional bosonic curved background.
The basic ingredients are:

i) a four-form potential $A_4(x)$ with curvature $F_5 = dA_4$;

ii) a metric $g_{mn}(x)$ (or the vielbeins $e^a = dx^m{e_m}^a$);

iii) an auxiliary scalar field $a(x)$ with its related one-form 
\beq
v = {{da}\over{\sqrt{-g^{mn}\partial_ma\partial_na}}} = e^av_a, 
\eeq 
normalized such that $ v^2 = v^a v_a = -1$.
Moreover, we associate to the anti--selfdual part of $F_5$, 
$F_5^{(-)} \equiv F_5 
- *F_5$, the four form obtained via the interior product with $v$:
$f_4 \equiv  i_v F_5^{(-)}$. 

The self duality condition, which has to be produced by the action, 
is $F_5^{(-)} = 0$; but, due to the identity 
\beq 
F_5^{(-)} = -vf_4 + *(vf_4),
\eeq 
$F_5^{(-)} = 0$ is equivalent to $f_4 = 0$.

The covariant action which, after gauge--fixing of new bosonic 
symmetries (see below),  leads to this equation is
\beq 
S_0 = \frac{1}{4} \int{(F_5 *F_5 +f_4 *f_4)}. 
\label{20}
\eeq 
It yields  the  field equations 
\beq 
d(da \tilde{f}_4) = 0, 
\label{21}
\eeq 
\beq 
d(da \tilde{f}_4\tilde{f}_4) = 0, 
\label{22}
\eeq
where $\tilde{f}_4$ = $f_4\over{\sqrt{- (\partial{a})^2}}$. 

In addition to the standard gauge trasformation for $A_4$, i.e.
$\delta A_4 = d\Lambda_3,$ $\delta a = 0$, the action (\ref{20}) 
is invariant under the new local 
symmetries 
\beq 
\delta_1 A_4 = -\phi \, \tilde{f}_4 ,\qquad \delta_1 a = \phi,
\label{23}
\eeq
\beq
\delta_2 A_4 = \psi_3 \, da, \qquad \delta_2 a = 0, 
\label{24} 
\eeq 
where $\phi(x)$ is an 
infinitesimal scalar but $\psi_3(x)$ is a finite 3-form. 
The symmetry (\ref{23}) implies that $a(x)$ is pure gauge (but the gauge 
$a = 0$ is forbidden, 
since $S_{0}$ is non polinomial in $a$ and becomes singular in $a=0$). 

The general solution of the field eq. (\ref{21}) is 
$vf_4 = d\tilde{\psi}_3da$,
and under the finite trasformation (\ref{24}) $vf_4$ trasforms as $vf_4 
\longmapsto vf_4 + d\psi_3 \, da$, so that, with this trasformation, 
one can reach the selfduality condition $f_4 =0$ i.e. 
$F_5^{(-)} = 0$.

This approach can be generalized to cover the case of a chiral boson 
$A_4$ interacting with matter fields, denoted collectively by $\chi$. 
The recipe is the following:

i) replace $F_5 = dA_4$ with $F_5 = dA_4 + \tilde{C}_5(x)$ where 
$\tilde{C}_5$ is a 5-form representing the coupling between $A_4$ and 
$\chi$;

ii) add the Wess-Zumino term 
\beq 
S_{WZ} = \frac{1}{2} \int \tilde{C}_5 dA_4 = \frac{1}{2} \int{F_5 dA_4};
\eeq

iii) add the action of $\chi$ in absence of $A_4$, $ S_{\chi} = 
 \int{L_{10}(\chi)}$, where $ L_{10}$ is a local $\chi$-dependent 
ten-form.

One can verify that the action 
\beq 
S=\int{\left(\, \frac{1}{4}\left(F_5 *F_5 + f_4 *f_4\right) + \frac{1}{2} 
F_5 \, dA_4 + L_{10}(\chi) \, \right)} 
\label{25} 
\eeq 
is invariant under the symmetries
(\ref{23}), (\ref{24})  and, after gauge fixing, yields the selfduality 
condition $F_5^{(-)} = 0$.

For $D=10$, $IIB$ SUGRA , the matter fields $\chi$ are the graviton, the 
gravitinos, the complex fermion $\Lambda$, the complex tensor $A_2$ and 
the scalars. According to (\ref{3}) and (\ref{8}), $\tilde{C}_5$ is 
\beq 
\tilde{C}_5 = i(A_{2I} \, d{A}^I_2 ) + C_5,
\eeq 
where $C_5$ is defined in eq. (\ref{12}). 

The problem is now to find $L_{10}$ such that the 
action S becomes invariant under $SU(1,1)$, under local $U(1)$ and local 
supersymmetry. Invariance under $SU(1,1)$ and $U(1)$ is automatic if $L_{10}$ 
is neutral with respect to $U(1)$ and contains only covariant derivatives and 
fields and forms which are scalars under $SU(1,1)$. Supersymmetry is 
more delicate. The susy transformations, as given by the superspace 
approach, close only on--shell, which in principle is not a problem.
But in the superspace approach "on shell" includes also the self--duality 
condition $F_5^{(-)} =0$ which, in our lagrangian approach, is not a field 
equation (it arises only after gauge fixing of the new symmetry (\ref{24})).
The problem concerns, actually, only the susy 
trasformation of the gravitinos which, as mentioned above, are the only 
ones which involve $F_5^{(+)}$,
the self--dual part of $F_5$. This means that the susy transformation of
the gravitino is determined only modulo terms which are proportional to the 
self--duality condition. Eventually these terms have to be fixed such that
the complete action becomes susy invariant. At first sight it is by no means 
obvious that such terms exist, but a detailed analysis leads to a quite
simple solution of this problem.
One  has to assume that the auxiliary 
field $a(x)$ is invariant under supersymmetry 
-- it is a non propagating field and has no fermionic partner -- 
and in the (on--shell) susy 
trasformation of the gravitinos, one has to replace $F_5^{(+)}$ with 
\beq 
K_5 = F_5 + vf_4. 
\label{26}
\eeq  
The five--form $K_5$ is an interesting object. 
It is selfdual, $K_5 =*K_5$, it coincides with $F_5^{(+)}$ if the selfduality 
condition (\ref{16}) holds, it is invariant under the tansformations 
(\ref{23}) and it is proportional to the field equation (\ref{21}) 
under (\ref{24}).

Now we can write the action for $D=10$, $IIB$ SUGRA:
\begin{eqnarray}
S &=& \int E_{ab} \, R^{ab} + \frac{1}{3} E_{abc} \, (i\bar{\psi}\Gamma^{abc}D\psi + 
c.c.) + 4 E_a \, (i\bar{\Lambda}\Gamma^aD\Lambda + c.c.) +
\nonumber \\
&+& \frac{1}{4} (F_5 * F_5 
+ f_4 * f_4) +\frac{1}{2} F_5 \, dA_4 -\frac{i}{2} (A_{2I} \, d{A}_2^I)\, C_5  +
\nonumber \\  
&+& 2\left[\bar{F}_3 * F_3 + \left(C_7 \, \bar{F}_3 - \frac{1}{2} C_7 \, \bar{C}_3 +c.c.\right)\right] + 
\nonumber \\
&+& 2\left[\bar{F}_1 * F_1 + \left(C_9 \, \bar{F}_1 - \frac{1}{2} C_9 \, \bar{C}_1 +c.c.\right)\right] +
\nonumber \\
&+& \left(\frac{1}{2} \bar{C}_7^{\psi}\, C_3^{\Lambda} - 2 
\bar{C}_7^{\Lambda} \, C_3^{\psi} + c.c.\right) + \frac{1}{2} \, C_5^{\Lambda} \, C_5^{\psi} -
3 E \, (\bar{\Lambda}\Gamma^a\Lambda) \, (\bar{\Lambda} \Gamma_a \Lambda), 
\label{27}
\end{eqnarray}
where 
\beq 
E_{{a_1}...{a_p}} = \frac{1}{(10-p)!} \epsilon_{{a_1}...{a_p}{b_1} 
...{b_{10-p}}}e^{b_1}...e^{b_{10-p}},
\eeq
and the $C_p$  are defined in (\ref{10})--(\ref{14}).
The first line of eq. (\ref{27}) contains the Einstein action and the kinetic 
terms for the fermions (gravitinos and $\Lambda$). The second line is 
the action for the self-dual tensor. The third and the fourth lines 
describe the kinetic actions for the rank-two tensors and for the 
scalars respectively. The four--fermions interactions are given in the 
last line. 

The covariant action S is manifestly invariant under 
$SL(2,R)$, the local  $U(1)$, the gauge symmetries of the potentials $A_2$ and 
$A_4$ and  under the new local symmetries (\ref{23}), (\ref{24}). It yields
the correct field equations and, after gauge fixing, the self--duality
 constraint for the chiral boson. It is also invariant under local 
supersymmetry, as it can be checked with a long calculation. The 
easiest way to do this is to lift to superspace all the forms which appear in 
S (except the term $f_4*f_4$), compute their covariant Lie derivative along 
$\epsilon$ and use the rheonomic parametrizations for the torsion and the 
curvatures. The term $f_4*f_4$ cannot be lifted to superspace since it contains 
the auxiliary field $a(x)$ which is not a superfield 
($\delta_{\epsilon}a(x) =0$) and its variation must be calculated by hand.
It is remarkable that under local supersymmetry transformations, 
the terms involving $A_4$ (the second line of (\ref{27})) transform 
in expressions containing $F_5$ and $v^a$ only in the combination (\ref{26}).
This variation can thus be cancelled 
by making the replacement $ {F_5}^{(+)} \mapsto K_5$ in the SUSY 
trasformation of the gravitinos.         
Let us  mention that, extending the method proposed in \cite{1}, one can 
also write actions \cite{4}, where the  basic fields, rank-two tensors and 
scalars and their duals, six-forms and eight-forms, appear in a symmetric 
way, as already shown in \cite{10} for the case of $D=11$ supergravity.
\vskip 1truecm
\paragraph{Acknowledgements.}
\ We are grateful to P. Pasti and D. Sorokin for their interest 
in this work and useful discussions. This work was supported by the 
European Commission TMR programme ERBFMPX-CT96-0045 to which K.L. and 
M.T. are associated.

%
%

\end{document}